\documentclass[10pt,a4paper,onecolumn]{article}
\usepackage[utf8]{inputenc}
\DeclareUnicodeCharacter{2207}{$\nabla$}
\usepackage{textcomp}
\usepackage{marginnote}
\usepackage{graphicx}
\usepackage{xcolor}
\usepackage{authblk,etoolbox}
\usepackage{titlesec}
\usepackage{calc}
\usepackage{tikz}
\usepackage{hyperref}
\hypersetup{colorlinks,breaklinks=true,
            urlcolor=[rgb]{0.0, 0.5, 1.0},
            linkcolor=[rgb]{0.0, 0.5, 1.0}}
\usepackage{caption}
\usepackage{tcolorbox}
\usepackage{amssymb,amsmath}
\usepackage{ifxetex,ifluatex}
\usepackage{seqsplit}
\usepackage{xstring}

\usepackage{float}
\let\origfigure\figure
\let\endorigfigure\endfigure
\renewenvironment{figure}[1][2] {
    \expandafter\origfigure\expandafter[H]
} {
    \endorigfigure
}

\usepackage{fixltx2e} 
\usepackage[
  backend=biber,
]{biblatex}
\bibliography{paper.bib}

\let\textttOrig=\texttt
\def\texttt#1{\expandafter\textttOrig{\seqsplit{#1}}}
\renewcommand{\seqinsert}{\ifmmode
  \allowbreak
  \else\penalty6000\hspace{0pt plus 0.02em}\fi}

\makeatletter
\let\href@Orig=\href
\def\href@Urllike#1#2{\href@Orig{#1}{\begingroup
    \def\Url@String{#2}\Url@FormatString
    \endgroup}}
\def\href@Notdoi#1#2{\def\tempa{#1}\def\tempb{#2}%
  \ifx\tempa\tempb\relax\href@Urllike{#1}{#2}\else
  \href@Orig{#1}{#2}\fi}
\def\href#1#2{%
  \IfBeginWith{#1}{https://doi.org}%
  {\href@Urllike{#1}{#2}}{\href@Notdoi{#1}{#2}}}
\makeatother

\newlength{\cslhangindent}
\newlength{\csllabelwidth}
\newenvironment{CSLReferences}[3] 
 {
  \setlength{\parindent}{0pt}
  \ifodd #1 \everypar{\setlength{\hangindent}{\cslhangindent}}\ignorespaces\fi
  \ifnum #2 > 0
  \setlength{\parskip}{#2\baselineskip}
  \fi
 }%
 {}
\usepackage{calc}

\usepackage[top=3.5cm, bottom=3cm, right=1.5cm, left=1.0cm,
            headheight=2.2cm, reversemp, includemp, marginparwidth=4.5cm]{geometry}

\titleformat{\section}
  {\normalfont\sffamily\Large\bfseries}
  {}{0pt}{}
\titleformat{\subsection}
  {\normalfont\sffamily\large\bfseries}
  {}{0pt}{}
\titleformat{\subsubsection}
  {\normalfont\sffamily\bfseries}
  {}{0pt}{}
\titleformat*{\paragraph}
  {\sffamily\normalsize}

\usepackage{fancyhdr}
\graphicspath{{figures/}{.}}
\makeatletter
\let\ps@plain\ps@fancy
\fancyheadoffset[L]{4.5cm}
\fancyfootoffset[L]{4.5cm}

\definecolor{linky}{rgb}{0.0, 0.5, 1.0}

\newtcolorbox{repobox}
   {colback=red, colframe=red!75!black,
     boxrule=0.5pt, arc=2pt, left=6pt, right=6pt, top=3pt, bottom=3pt}

\newcommand{\ExternalLink}{%
   \tikz[x=1.2ex, y=1.2ex, baseline=-0.05ex]{%
       \begin{scope}[x=1ex, y=1ex]
           \clip (-0.1,-0.1)
               --++ (-0, 1.2)
               --++ (0.6, 0)
               --++ (0, -0.6)
               --++ (0.6, 0)
               --++ (0, -1);
           \path[draw,
               line width = 0.5,
               rounded corners=0.5]
               (0,0) rectangle (1,1);
       \end{scope}
       \path[draw, line width = 0.5] (0.5, 0.5)
           -- (1, 1);
       \path[draw, line width = 0.5] (0.6, 1)
           -- (1, 1) -- (1, 0.6);
       }
   }

\patchcmd{\@maketitle}{center}{flushleft}{}{}
\patchcmd{\@maketitle}{center}{flushleft}{}{}
\patchcmd{\@maketitle}{\LARGE}{\LARGE\sffamily}{}{}
\def\maketitle{{%
  
  \AB@maketitle}}
\makeatletter
\renewcommand\AB@affilsepx{ \protect\Affilfont}
\renewcommand\AB@affilnote[1]{{\bfseries #1}\hspace{3pt}}
\renewcommand{\affil}[2][]%
   {\newaffiltrue\let\AB@blk@and\AB@pand
      \if\relax#1\relax\def\AB@note{\AB@thenote}\else\def\AB@note{#1}%
        \setcounter{Maxaffil}{0}\fi
        \begingroup
        \let\href=\href@Orig
        \let\texttt=\textttOrig
        \let\protect\@unexpandable@protect
        \def\thanks{\protect\thanks}\def\footnote{\protect\footnote}%
        \@temptokena=\expandafter{\AB@authors}%
        {\def\\{\protect\\\protect\Affilfont}\xdef\AB@temp{#2}}%
         \xdef\AB@authors{\the\@temptokena\AB@las\AB@au@str
         \protect\\[\affilsep]\protect\Affilfont\AB@temp}%
         \gdef\AB@las{}\gdef\AB@au@str{}%
        {\def\\{, \ignorespaces}\xdef\AB@temp{#2}}%
        \@temptokena=\expandafter{\AB@affillist}%
        \xdef\AB@affillist{\the\@temptokena \AB@affilsep
          \AB@affilnote{\AB@note}\protect\Affilfont\AB@temp}%
      \endgroup
       \let\AB@affilsep\AB@affilsepx
}
\makeatother

\renewcommand\Affilfont{\sffamily\small\mdseries}
\ifnum 0\ifxetex 1\fi\ifluatex 1\fi=0 
  \usepackage[T1]{fontenc}
  \usepackage[utf8]{inputenc}

\else 
  \ifxetex
    \usepackage{mathspec}
    \usepackage{fontspec}

  \else
    \usepackage{fontspec}
  \fi
  \defaultfontfeatures{Ligatures=TeX,Scale=MatchLowercase}

\fi
\IfFileExists{upquote.sty}{\usepackage{upquote}}{}
\IfFileExists{microtype.sty}{%
\usepackage{microtype}
\UseMicrotypeSet[protrusion]{basicmath} 
}{}

\usepackage{hyperref}
\hypersetup{unicode=true,
            pdftitle={iharm3D: Vectorized General Relativistic Magnetohydrodynamics},
            pdfborder={0 0 0},
            breaklinks=true}
\let\addcontentslineOrig=\addcontentsline
\def\addcontentsline#1#2#3{\bgroup
  \let\texttt=\textttOrig\addcontentslineOrig{#1}{#2}{#3}\egroup}
\let\markbothOrig\markboth
\def\markboth#1#2{\bgroup
  \let\texttt=\textttOrig\markbothOrig{#1}{#2}\egroup}
\let\markrightOrig\markright
\def\markright#1{\bgroup
  \let\texttt=\textttOrig\markrightOrig{#1}\egroup}

\usepackage{graphicx,grffile}
\makeatletter
\def\maxwidth{\ifdim\Gin@nat@width>\linewidth\linewidth\else\Gin@nat@width\fi}
\def\maxheight{\ifdim\Gin@nat@height>\textheight\textheight\else\Gin@nat@height\fi}
\makeatother
\setkeys{Gin}{width=\maxwidth,height=\maxheight,keepaspectratio}
\IfFileExists{parskip.sty}{%
\usepackage{parskip}
}{
\setlength{\parindent}{0pt}
\setlength{\parskip}{6pt plus 2pt minus 1pt}
}
\providecommand{\tightlist}{%
  \setlength{\itemsep}{0pt}\setlength{\parskip}{0pt}}
\ifx\paragraph\undefined\else
\let\oldparagraph\paragraph
\renewcommand{\paragraph}[1]{\oldparagraph{#1}\mbox{}}
\fi
\ifx\subparagraph\undefined\else
\let\oldsubparagraph\subparagraph
\renewcommand{\subparagraph}[1]{\oldsubparagraph{#1}\mbox{}}
\fi

\title{iharm3D: Vectorized General Relativistic Magnetohydrodynamics}

        \author[1, 2]{Cora Prather\footnote{Corresponding author}}
          \author[1, 2]{George N. Wong}
          \author[1, 2]{Vedant Dhruv}
          \author[3]{Benjamin R. Ryan}
          \author[3]{Joshua C. Dolence}
          \author[5]{Sean M. Ressler}
          \author[1, 2, 4]{Charles F. Gammie}
    
      \affil[1]{Physics Department, University of Illinois at
Urbana--Champaign, 1110 West Green Street, Urbana, IL 61801, USA}
      \affil[2]{Illinois Center for Advanced Studies of the Universe}
      \affil[3]{CCS-2, Los Alamos National Laboratory, P.O. Box 1663,
Los Alamos, NM 87545, USA}
      \affil[4]{Astronomy Department, University of Illinois at
Urbana--Champaign, 1002 West Green Street, Urbana, IL 61801, USA}
      \affil[5]{Kavli Institute for Theoretical Physics, University of
California Santa Barbara, Kohn Hall, Santa Barbara, CA 93107, USA}
  \date{\vspace{-7ex}}

\begin{document}
\maketitle

\marginpar{

  \begin{flushleft}
  \sffamily\small

  {\bfseries DOI:} \href{https://doi.org/DOI unavailable}{\color{linky}{DOI unavailable}}

  \vspace{2mm}

  {\bfseries Software}
  \begin{itemize}
    \setlength\itemsep{0em}
    \item \href{N/A}{\color{linky}{Review}} \ExternalLink
    \item \href{NO_REPOSITORY}{\color{linky}{Repository}} \ExternalLink
    \item \href{DOI unavailable}{\color{linky}{Archive}} \ExternalLink
  \end{itemize}

  \vspace{2mm}

  \par\noindent\hrulefill\par

  \vspace{2mm}

  {\bfseries Editor:} \href{https://example.com}{Pending
Editor} \ExternalLink \\
  \vspace{1mm}
    {\bfseries Reviewers:}
  \begin{itemize}
  \setlength\itemsep{0em}
    \item \href{https://github.com/Pending Reviewers}{@Pending
Reviewers}
    \end{itemize}
    \vspace{2mm}

  {\bfseries Submitted:} N/A\\
  {\bfseries Published:} N/A

  \vspace{2mm}
  {\bfseries License}\\
  Authors of papers retain copyright and release the work under a Creative Commons Attribution 4.0 International License (\href{http://creativecommons.org/licenses/by/4.0/}{\color{linky}{CC BY 4.0}}).

  \end{flushleft}
}

\hypertarget{iharm3d-functionality-and-purpose}{%
\section{\texorpdfstring{\texttt{iharm3D} Functionality and
Purpose}{iharm3D Functionality and Purpose}}\label{iharm3d-functionality-and-purpose}}

\texttt{iharm3D}\footnote{https://github.com/AFD-Illinois/iharm3d} is an
open-source C code for simulating black hole accretion systems in
arbitrary stationary spacetimes using ideal general-relativistic
magnetohydrodynamics (GRMHD). It is an implementation of the HARM
(``High Accuracy Relativistic Magnetohydrodynamics'') algorithm outlined
in Gammie et al. (2003) with updates as outlined in McKinney \& Gammie
(2004) and Noble et al. (2006). The code is most directly derived from
Ryan et al. (2015) but with radiative transfer portions removed. HARM is
a conservative finite-volume scheme for solving the equations of ideal
GRMHD, a hyperbolic system of partial differential equations, on a
logically Cartesian mesh in arbitrary coordinates.

\hypertarget{statement-of-need}{%
\section{Statement of Need}\label{statement-of-need}}

Numerical simulations are crucial in modeling observations of active
galactic nuclei, such as the recent horizon-scale results from the Event
Horizon Telescope and GRAVITY collaborations. The computational
simplicity of ideal GRMHD enables the generation of long,
high-resolution simulations and broad parameter-exploration studies that
can be compared to observations for parameter inference.

Multiple codes already exist for solving the ideal GRMHD equations on
regular Eulerian meshes in 3D, including:

\begin{itemize}
\tightlist
\item
  Athena++ (Stone et al. (2020), White et al. (2016))
\item
  BHAC (Porth et al. (2017))
\item
  Cosmos++ (Anninos et al. (2005), Fragile et al. (2012), Fragile et al.
  (2014))
\item
  ECHO (Londrillo \& Zanna (2000), Londrillo \& Zanna (2004))
\item
  H-AMR (Matthew Liska et al. (2019), M. Liska et al. (2020))
\item
  HARM-Noble (Noble et al. (2006), Noble et al. (2009), Noble et al.
  (2012), Zilhão \& Noble (2014), Bowen et al. (2018))
\item
  IllinoisGRMHD (Etienne et al. (2015))
\item
  KORAL (Sądowski et al. (2013), Sądowski et al. (2014))
\item
  GRHydro (Mösta et al. (2014))
\item
  Spritz (Cipolletta et al. (2020), Cipolletta et al. (2021))
\end{itemize}

As the length of this list illustrates, the field of GRMHD simulation is
now well established, and many codes now exist to serve different needs.
These codes can be distinguished by the trade-offs they make in
prioritizing speed, simplicity, and generality, with the latter
encompassing, e.g., support for dynamical spacetimes, adaptive mesh
refinement, or higher-order integration schemes.

In particular, \texttt{iharm3D} aims to be a simple and fast code
capable of simulating the original systems of interest when designing
HARM, even at the cost of features aimed at more general applicability.
It provides a fast and scalable update to HARM, but maintains the
conventions and structure of the original described in Gammie et al.
(2003). The result is a code that is relatively easy to understand and
modify, yet capable of running simulations at state-of-the-art scale.

\hypertarget{implementation-notes}{%
\section{Implementation Notes}\label{implementation-notes}}

In MHD, uncorrected discretization errors inevitably lead to violations
of the no-monopoles condition \(\nabla \cdot B = 0\). As in the original
HARM implementation, \texttt{iharm3D} uses the ``Flux-CT'' scheme for
cell-centered constrained transport outlined in Tóth (2000).

\texttt{iharm3D} also retains numerical evaluation of all
metric-dependent quantities, allowing trivial modification of the
coordinate system or background spacetime so long as the line element is
available in analytic form. This can be used as a form of static mesh
refinement, since the coordinates can be adapted to place resolution in
areas of interest (e.g., near the accretion disk midplane).

In GRMHD, ``conserved'' variables (energy and momentum densities) are
complicated analytic functions of ``primitive'' variables (density,
pressure, and velocity). Conserved variables are stepped forward in time
and then inversion to primitives is done numerically. \texttt{iharm3D}
uses the ``\(1D_W\)'' scheme outlined in Noble et al. (2006).

As the equations of ideal GRMHD are rescalable, any consistent set of
units may be chosen to evolve them in the code. For numerical stability,
when simulating accretion systems we choose units in which
\(GM = c = 1\) with \(M\) the mass of the central object, and scale the
density of the initial conditions such that the maximum value of
\(\rho\) is \(1\).

To model a collisionless plasma, \texttt{iharm3D} implements an optional
means of tracking and partitioning dissipation into ions and electrons
(Ressler et al. (2015)). Currently the code implements five different
heating models, described in Howes (2010), Kawazura et al. (2019),
Werner et al. (2018), Rowan et al. (2017), and Sharma et al. (2007).

To avoid catastrophic failures caused by discretization error,
especially in low density regions, fluid variables are bounded at the
end of each step. Typically, the bounds in black hole accretion problems
are enforced as follows:

\begin{itemize}
\tightlist
\item
  Density \(\rho > 10^{-6} k\), for
  \(k \equiv \frac{1}{r^2 (1 + r/10)}\), with \(r\) the radial
  coordinate,
\item
  Internal energy density \(u > 10^{-8} k^{\gamma}\) where
  \(\gamma \equiv\) adiabatic index,
\item
  \(\rho\) and \(u\) are incremented until
  \(\sigma \equiv \frac{2 P_b}{\rho} < 400\) and
  \(\beta \equiv \frac{P_{gas}}{P_b} > 2.5 \times 10^{-5}\) where
  \(P_b \equiv \frac{b^2}{2}\) is the magnetic pressure,
\item
  \(\rho\) is incremented until \(\frac{u}{\rho} < 100\),
\item
  When evolving electron temperatures, \(u\) is decremented until
  \(\frac{P_{gas}}{\rho^{\gamma}} < 3\),
\item
  Velocity components are downscaled until Lorentz factor
  \(\Gamma \equiv \frac{1}{\sqrt{1 - v^2}} < 50\).
\end{itemize}

Global disk simulations inevitably invoke these bounds, most frequently
those on \(\sigma\) and \(\Gamma\).

\hypertarget{tests}{%
\section{Tests}\label{tests}}

\begin{figure}
\centering
\includegraphics[width=5in,height=\textheight]{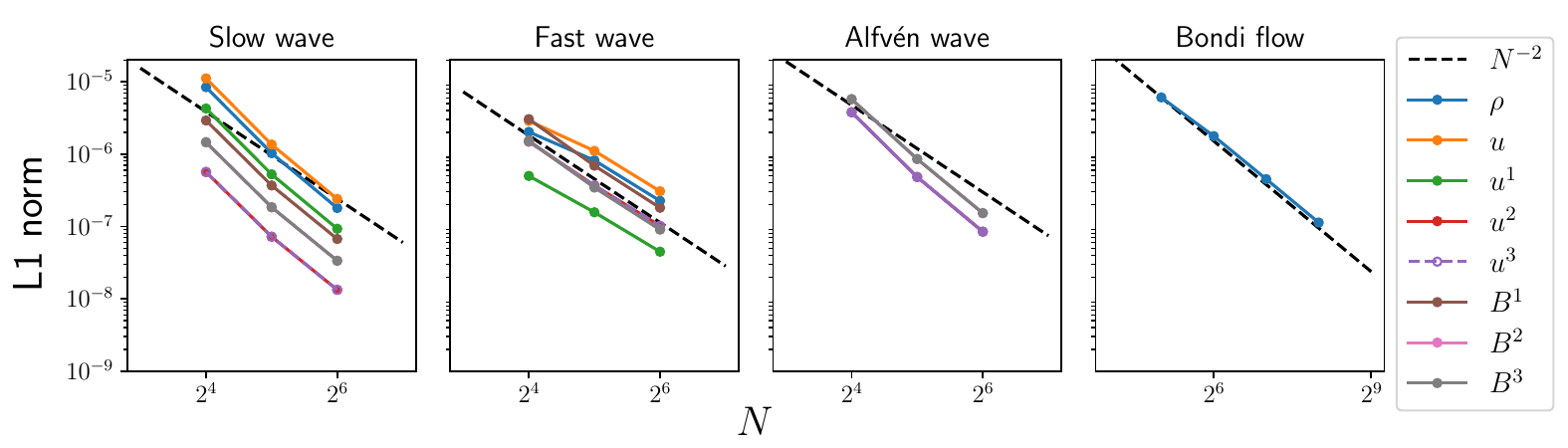}
\caption{Results of convergence tests with \texttt{iharm3D}'s main
branch, plotting L1 norm of the difference between the computed solution
and the analytic or stable result with increasing domain size. Wave
solutions were performed on a 3D cubic grid \(N\) zones to one side, the
Bondi accretion problem was performed on a logically Cartesian 2D square
grid \(N\) zones on one side. \label{fig:convergence}}
\end{figure}

The convergence properties of HARM are well-studied in Gammie et al.
(2003). \texttt{iharm3D} implements most of the tests presented in that
paper as integration and regression tests. Figure \ref{fig:convergence}
shows convergence results for linear modes and for un-magnetized Bondi
flow.

\texttt{iharm3D} implements three additional tests which check that
fluid evolution is identical for different domain decompositions: one
which initializes a new fluid state, one which restarts from a
checkpoint file, and one comparing the initialized state to an
equivalent checkpoint file.

\hypertarget{scaling}{%
\section{Scaling}\label{scaling}}

\begin{figure}
\centering
\includegraphics[width=4.5in,height=\textheight]{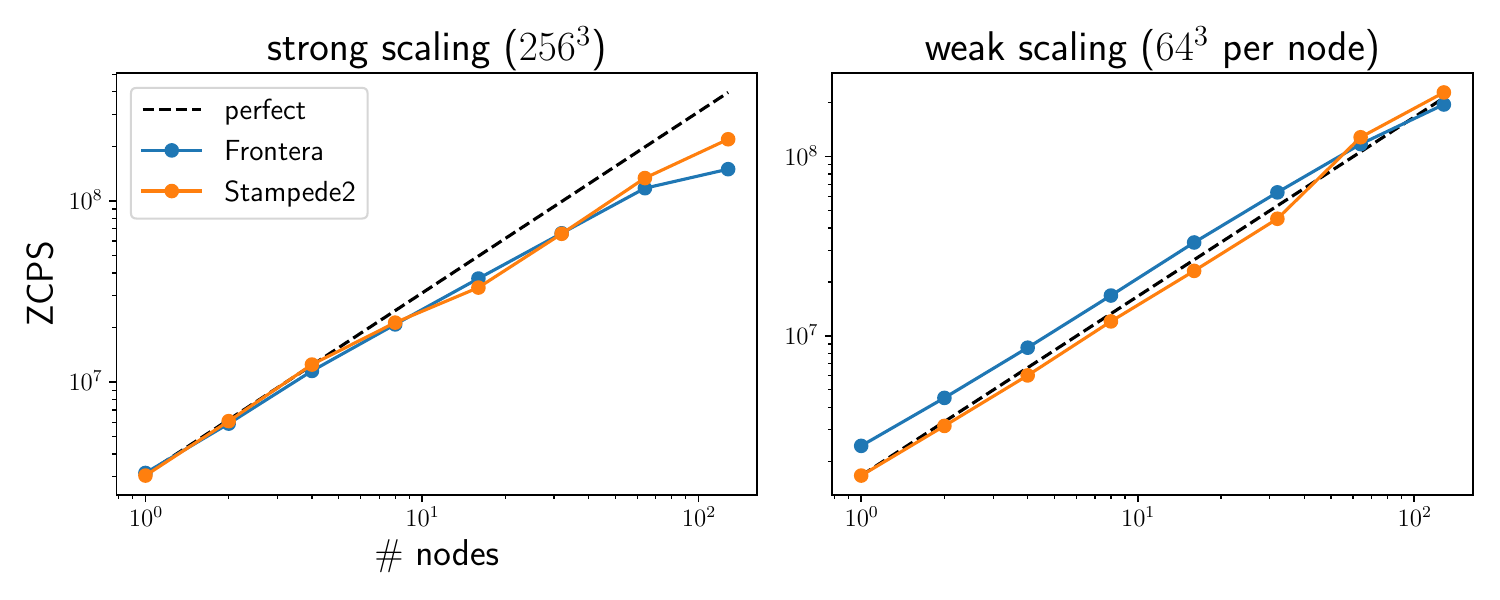}
\caption{Strong and weak scaling performance of iharm3D. Performance is
measured in zones advanced by one cycle each second (Zone-Cycles per
Second). In the strong scaling test, an accretion torus problem of
constant total size \(256^3\) was split among all nodes; in the weak
scaling test, the total problem size was varied to ensure that a mesh
block of size \(64^3\) was allocated to each node. \label{fig:scaling}}
\end{figure}

Key \texttt{iharm3D} routines are written for effective compiler
vectorization and prioritize simple memory access patterns to make good
use of high memory bandwidth. Originally developed for Intel Knights
Landing (KNL) chips on the Stampede2 supercomputer at Texas Advanced
Computing Center (TACC), \texttt{iharm3D} also runs efficiently on TACC
Frontera, which uses traditional Cascade Lake (CLX) CPUs. Figure
\ref{fig:scaling} presents scaling results for \texttt{iharm3D} on both
Stampede2 and Frontera.

\hypertarget{research-projects-using-iharm3d}{%
\section{Research projects using
iharm3D}\label{research-projects-using-iharm3d}}

\texttt{iharm3D} is one of several GRMHD codes used by the EHT
Collaboration to produce its library of fluid simulations. Images
produced from this library were used for validation tests in Event
Horizon Telescope Collaboration et. al. (2019a) and Event Horizon
Telescope Collaboration et. al. (2021a) and for interpretation of the
M87 EHT results in total intensity (Event Horizon Telescope
Collaboration et. al. (2019b), Event Horizon Telescope Collaboration et.
al. (2019c)) and polarization (Event Horizon Telescope Collaboration et.
al. (2021b)).

Papers making use of the results of \texttt{iharm3D} simulations include
Porth et al. (2019), Johnson et al. (2020), Gold et al. (2020), Palumbo
et al. (2020), Lin et al. (2020), Ricarte et al. (2020), Wielgus et al.
(2020), Tiede et al. (2020), and Gelles et al. (2021).

\hypertarget{acknowledgements}{%
\section{Acknowledgements}\label{acknowledgements}}

This work was supported by National Science Foundation grants AST
17-16327, OISE 17-43747, AST 20-07936, AST 20-34306, and PHY 17-48958,
by a Donald C. and F. Shirley Jones Fellowship to G.N.W., by the Gordon
and Betty Moore Foundation through Grant GBMF7392, and by the US
Department of Energy through Los Alamos National Laboratory. Los Alamos
National Laboratory is operated by Triad National Security, LLC, for the
National Nuclear Security Administration of the US Department of Energy
(Contract No.~89233218CNA000001). This work has been assigned a document
release number LA-UR-21-23714.

This work used the Extreme Science and Engineering Discovery Environment
(XSEDE), which is supported by National Science Foundation grant number
ACI-1548562, specifically the XSEDE resource Stampede2 at the Texas
Advanced Computing Center (TACC) through allocation TG-AST170024. The
authors acknowledge the Texas Advanced Computing Center at The
University of Texas at Austin for providing HPC resources that have
contributed to the research results reported within this paper.

\hypertarget{references}{%
\section*{References}\label{references}}
\addcontentsline{toc}{section}{References}

\hypertarget{refs}{}
\begin{CSLReferences}{1}{0}
\leavevmode\hypertarget{ref-anninos_cosmos_2005}{}%
Anninos, P., Fragile, P. C., \& Salmonson, J. D. (2005). Cosmos++:
{Relativistic} {Magnetohydrodynamics} on {Unstructured} {Grids} with
{Local} {Adaptive} {Refinement}. \emph{The Astrophysical Journal},
\emph{635}(1), 723. \url{https://doi.org/10.1086/497294}

\leavevmode\hypertarget{ref-bowen_quasi-periodic_2018}{}%
Bowen, D. B., Mewes, V., Campanelli, M., Noble, S. C., Krolik, J. H., \&
Zilhão, M. (2018). Quasi-periodic {Behavior} of {Mini}-disks in {Binary}
{Black} {Holes} {Approaching} {Merger}. \emph{The Astrophysical Journal
Letters}, \emph{853}(1), L17.
\url{https://doi.org/10.3847/2041-8213/aaa756}

\leavevmode\hypertarget{ref-cipolletta_spritz_2020}{}%
Cipolletta, F., Kalinani, J. V., Giacomazzo, B., \& Ciolfi, R. (2020).
Spritz: A new fully general-relativistic magnetohydrodynamic code.
\emph{Classical and Quantum Gravity}, \emph{37}, 135010.
\url{https://doi.org/10.1088/1361-6382/ab8be8}

\leavevmode\hypertarget{ref-cipolletta_spritz_2021}{}%
Cipolletta, F., Kalinani, J. V., Giangrandi, E., Giacomazzo, B., Ciolfi,
R., Sala, L., \& Giudici, B. (2021). Spritz: General relativistic
magnetohydrodynamics with neutrinos. \emph{Classical and Quantum
Gravity}, \emph{38}, 085021.
\url{https://doi.org/10.1088/1361-6382/abebb7}

\leavevmode\hypertarget{ref-etienne_illinoisgrmhd_2015}{}%
Etienne, Z. B., Paschalidis, V., Haas, R., Mösta, P., \& Shapiro, S. L.
(2015). {IllinoisGRMHD}: An open-source, user-friendly {GRMHD} code for
dynamical spacetimes. \emph{Classical and Quantum Gravity},
\emph{32}(17), 175009.
\url{https://doi.org/10.1088/0264-9381/32/17/175009}

\leavevmode\hypertarget{ref-PaperVII}{}%
Event Horizon Telescope Collaboration et. al. (2021a). {First M87 Event
Horizon Telescope Results. VII. Polarization of the Ring}.
\emph{910}(1), L12. \url{https://doi.org/10.3847/2041-8213/abe71d}

\leavevmode\hypertarget{ref-PaperVIII}{}%
Event Horizon Telescope Collaboration et. al. (2021b). {First M87 Event
Horizon Telescope Results. VIII. Magnetic Field Structure near The Event
Horizon}. \emph{910}(1), L13.
\url{https://doi.org/10.3847/2041-8213/abe4de}

\leavevmode\hypertarget{ref-PaperIV}{}%
Event Horizon Telescope Collaboration et. al. (2019a). {First M87 Event
Horizon Telescope Results. IV. Imaging the Central Supermassive Black
Hole}. \emph{875}(1), L4. \url{https://doi.org/10.3847/2041-8213/ab0e85}

\leavevmode\hypertarget{ref-PaperV}{}%
Event Horizon Telescope Collaboration et. al. (2019b). {First M87 Event
Horizon Telescope Results. V. Physical Origin of the Asymmetric Ring}.
\emph{875}(1), L5. \url{https://doi.org/10.3847/2041-8213/ab0f43}

\leavevmode\hypertarget{ref-PaperVI}{}%
Event Horizon Telescope Collaboration et. al. (2019c). {First M87 Event
Horizon Telescope Results. VI. The Shadow and Mass of the Central Black
Hole}. \emph{875}(1), L6. \url{https://doi.org/10.3847/2041-8213/ab1141}

\leavevmode\hypertarget{ref-fragile_numerical_2012}{}%
Fragile, P. C., Gillespie, A., Monahan, T., Rodriguez, M., \& Anninos,
P. (2012). Numerical {Simulations} of {Optically} {Thick} {Accretion}
onto a {Black} {Hole}. {I}. {Spherical} {Case}. \emph{The Astrophysical
Journal Supplement Series}, \emph{201}, 9.
\url{https://doi.org/10.1088/0067-0049/201/2/9}

\leavevmode\hypertarget{ref-fragile_numerical_2014}{}%
Fragile, P. C., Olejar, A., \& Anninos, P. (2014). Numerical
{Simulations} of {Optically} {Thick} {Accretion} onto a {Black} {Hole}.
{II}. {Rotating} {Flow}. \emph{The Astrophysical Journal}, \emph{796},
22. \url{https://doi.org/10.1088/0004-637X/796/1/22}

\leavevmode\hypertarget{ref-gammie_harm:_2003}{}%
Gammie, C. F., McKinney, J. C., \& Tóth, G. (2003). {HARM}: {A}
{Numerical} {Scheme} for {General} {Relativistic}
{Magnetohydrodynamics}. \emph{The Astrophysical Journal}, \emph{589}(1),
444. \url{https://doi.org/10.1086/374594}

\leavevmode\hypertarget{ref-gelles_role_2021}{}%
Gelles, Z., Prather, B. S., Palumbo, D. C. M., Johnson, M. D., Wong, G.
N., \& Georgiev, B. (2021). The {Role} of {Adaptive} {Ray} {Tracing} in
{Analyzing} {Black} {Hole} {Structure}. \emph{The Astrophysical
Journal}, \emph{912}(1), 39.
\url{https://doi.org/10.3847/1538-4357/abee13}

\leavevmode\hypertarget{ref-gold2020}{}%
Gold, R., Broderick, A. E., Younsi, Z., Fromm, C. M., Gammie, C. F.,
Mościbrodzka, M., Pu, H.-Y., Bronzwaer, T., Davelaar, J., Dexter, J.,
Ball, D., Chan, C., Kawashima, T., Mizuno, Y., Ripperda, B., Akiyama,
K., Alberdi, A., Alef, W., Asada, K., \ldots{} Event Horizon Telescope
Collaboration. (2020). {Verification of Radiative Transfer Schemes for
the EHT}. \emph{897}(2), 148.
\url{https://doi.org/10.3847/1538-4357/ab96c6}

\leavevmode\hypertarget{ref-howes_prescription_2010}{}%
Howes, G. G. (2010). A prescription for the turbulent heating of
astrophysical plasmas. \emph{Monthly Notices of the Royal Astronomical
Society}, \emph{409}, L104--L108.
\url{https://doi.org/10.1111/j.1745-3933.2010.00958.x}

\leavevmode\hypertarget{ref-johnson_universal_2020}{}%
Johnson, M. D., Lupsasca, A., Strominger, A., Wong, G. N., Hadar, S.,
Kapec, D., Narayan, R., Chael, A., Gammie, C. F., Galison, P., Palumbo,
D. C. M., Doeleman, S. S., Blackburn, L., Wielgus, M., Pesce, D. W.,
Farah, J. R., \& Moran, J. M. (2020). Universal interferometric
signatures of a black hole's photon ring. \emph{Science Advances},
\emph{6}(12), eaaz1310. \url{https://doi.org/10.1126/sciadv.aaz1310}

\leavevmode\hypertarget{ref-kawazura_thermal_2019}{}%
Kawazura, Y., Barnes, M., \& Schekochihin, A. A. (2019). Thermal
disequilibration of ions and electrons by collisionless plasma
turbulence. \emph{Proceedings of the National Academy of Science},
\emph{116}, 771--776. \url{https://doi.org/10.1073/pnas.1812491116}

\leavevmode\hypertarget{ref-lin_feature_2020}{}%
Lin, J. Y.-Y., Wong, G. N., Prather, B. S., \& Gammie, C. F. (2020).
Feature {Extraction} on {Synthetic} {Black} {Hole} {Images}.
\emph{arXiv:2007.00794 {[}Astro-Ph{]}}.
\url{http://arxiv.org/abs/2007.00794}

\leavevmode\hypertarget{ref-liska_h-amr_2019}{}%
Liska, Matthew, Chatterjee, K., Tchekhovskoy, A., Yoon, D., Eijnatten,
D. van, Hesp, C., Markoff, S., Ingram, A., \& Klis, M. van der. (2019).
H-{AMR}: {A} {New} {GPU}-accelerated {GRMHD} {Code} for {Exascale}
{Computing} {With} {3D} {Adaptive} {Mesh} {Refinement} and {Local}
{Adaptive} {Time}-stepping. \emph{arXiv:1912.10192 {[}Astro-Ph{]}}.
\url{http://arxiv.org/abs/1912.10192}

\leavevmode\hypertarget{ref-liska_large-scale_2020}{}%
Liska, M., Tchekhovskoy, A., \& Quataert, E. (2020). Large-scale
poloidal magnetic field dynamo leads to powerful jets in {GRMHD}
simulations of black hole accretion with toroidal field. \emph{Monthly
Notices of the Royal Astronomical Society}, \emph{494}, 3656--3662.
\url{https://doi.org/10.1093/mnras/staa955}

\leavevmode\hypertarget{ref-londrillo_high-order_2000}{}%
Londrillo, P., \& Zanna, L. D. (2000). High-{Order} {Upwind} {Schemes}
for {Multidimensional} {Magnetohydrodynamics}. \emph{The Astrophysical
Journal}, \emph{530}(1), 508. \url{https://doi.org/10.1086/308344}

\leavevmode\hypertarget{ref-londrillo_divergence-free_2004}{}%
Londrillo, P., \& Zanna, L. del. (2004). On the divergence-free
condition in {Godunov}-type schemes for ideal magnetohydrodynamics: The
upwind constrained transport method. \emph{Journal of Computational
Physics}, \emph{195}, 17--48.
\url{https://doi.org/10.1016/j.jcp.2003.09.016}

\leavevmode\hypertarget{ref-mckinney_measurement_2004}{}%
McKinney, J. C., \& Gammie, C. F. (2004). A {Measurement} of the
{Electromagnetic} {Luminosity} of a {Kerr} {Black} {Hole}. \emph{The
Astrophysical Journal}, \emph{611}(2), 977.
\url{https://doi.org/10.1086/422244}

\leavevmode\hypertarget{ref-mosta_grhydro_2014}{}%
Mösta, P., Mundim, B. C., Faber, J. A., Haas, R., Noble, S. C., Bode,
T., Löffler, F., Ott, C. D., Reisswig, C., \& Schnetter, E. (2014).
{GRHydro}: A new open-source general-relativistic magnetohydrodynamics
code for the {Einstein} toolkit. \emph{Classical and Quantum Gravity},
\emph{31}, 015005. \url{https://doi.org/10.1088/0264-9381/31/1/015005}

\leavevmode\hypertarget{ref-noble_primitive_2006}{}%
Noble, S. C., Gammie, C. F., McKinney, J. C., \& Del Zanna, L. (2006).
Primitive {Variable} {Solvers} for {Conservative} {General}
{Relativistic} {Magnetohydrodynamics}. \emph{The Astrophysical Journal},
\emph{641}, 626--637. \url{https://doi.org/10.1086/500349}

\leavevmode\hypertarget{ref-noble_direct_2009}{}%
Noble, S. C., Krolik, J. H., \& Hawley, J. F. (2009). {DIRECT}
{CALCULATION} {OF} {THE} {RADIATIVE} {EFFICIENCY} {OF} {AN} {ACCRETION}
{DISK} {AROUND} {A} {BLACK} {HOLE}. \emph{The Astrophysical Journal},
\emph{692}(1), 411--421.
\url{https://doi.org/10.1088/0004-637X/692/1/411}

\leavevmode\hypertarget{ref-noble_circumbinary_2012}{}%
Noble, S. C., Mundim, B. C., Nakano, H., Krolik, J. H., Campanelli, M.,
Zlochower, Y., \& Yunes, N. (2012). Circumbinary {Magnetohydrodynamic}
{Accretion} into {Inspiraling} {Binary} {Black} {Holes}. \emph{The
Astrophysical Journal}, \emph{755}, 51.
\url{https://doi.org/10.1088/0004-637X/755/1/51}

\leavevmode\hypertarget{ref-palumbo_discriminating_2020}{}%
Palumbo, D. C. M., Wong, G. N., \& Prather, B. S. (2020). Discriminating
{Accretion} {States} via {Rotational} {Symmetry} in {Simulated}
{Polarimetric} {Images} of {M87}. \emph{The Astrophysical Journal},
\emph{894}(2), 156. \url{https://doi.org/10.3847/1538-4357/ab86ac}

\leavevmode\hypertarget{ref-porth_event_2019}{}%
Porth, O., Chatterjee, K., Narayan, R., Gammie, C. F., Mizuno, Y.,
Anninos, P., Baker, J. G., Bugli, M., Chan, C., Davelaar, J., Del Zanna,
L., Etienne, Z. B., Fragile, P. C., Kelly, B. J., Liska, M., Markoff,
S., McKinney, J. C., Mishra, B., Noble, S. C., \ldots{} Collaboration,
T. E. H. T. (2019). The {Event} {Horizon} {General} {Relativistic}
{Magnetohydrodynamic} {Code} {Comparison} {Project}.
\emph{arXiv:1904.04923 {[}Astro-Ph, Physics:gr-Qc{]}}.
\url{https://doi.org/10.3847/1538-4365/ab29fd}

\leavevmode\hypertarget{ref-porth_black_2017}{}%
Porth, O., Olivares, H., Mizuno, Y., Younsi, Z., Rezzolla, L.,
Moscibrodzka, M., Falcke, H., \& Kramer, M. (2017). The {Black} {Hole}
{Accretion} {Code}. \emph{Computational Astrophysics and Cosmology},
\emph{4}(1). \url{https://doi.org/10.1186/s40668-017-0020-2}

\leavevmode\hypertarget{ref-ressler_electron_2015}{}%
Ressler, S. M., Tchekhovskoy, A., Quataert, E., Chandra, M., \& Gammie,
C. F. (2015). Electron thermodynamics in {GRMHD} simulations of
low-luminosity black hole accretion. \emph{Monthly Notices of the Royal
Astronomical Society}, \emph{454}, 1848--1870.
\url{https://doi.org/10.1093/mnras/stv2084}

\leavevmode\hypertarget{ref-ricarte_decomposing_2020}{}%
Ricarte, A., Prather, B. S., Wong, G. N., Narayan, R., Gammie, C., \&
Johnson, M. D. (2020). Decomposing the internal faraday rotation of
black hole accretion flows. \emph{Monthly Notices of the Royal
Astronomical Society}, \emph{498}, 5468--5488.
\url{https://doi.org/10.1093/mnras/staa2692}

\leavevmode\hypertarget{ref-rowan_electron_2017}{}%
Rowan, M. E., Sironi, L., \& Narayan, R. (2017). Electron and {Proton}
{Heating} in {Transrelativistic} {Magnetic} {Reconnection}. \emph{The
Astrophysical Journal}, \emph{850}, 29.
\url{https://doi.org/10.3847/1538-4357/aa9380}

\leavevmode\hypertarget{ref-ryan_bhlight:_2015}{}%
Ryan, B. R., Dolence, J. C., \& Gammie, C. F. (2015). Bhlight: {General}
{Relativistic} {Radiation} {Magnetohydrodynamics} with {Monte} {Carlo}
{Transport}. \emph{The Astrophysical Journal}, \emph{807}(1), 31.
\url{https://doi.org/10.1088/0004-637X/807/1/31}

\leavevmode\hypertarget{ref-sadowski_numerical_2014}{}%
Sądowski, A., Narayan, R., McKinney, J. C., \& Tchekhovskoy, A. (2014).
Numerical simulations of super-critical black hole accretion flows in
general relativity. \emph{Monthly Notices of the Royal Astronomical
Society}, \emph{439}(1), 503.
\url{https://doi.org/10.1093/mnras/stt2479}

\leavevmode\hypertarget{ref-sadowski_semi-implicit_2013}{}%
Sądowski, A., Narayan, R., Tchekhovskoy, A., \& Zhu, Y. (2013).
Semi-implicit scheme for treating radiation under {M1} closure in
general relativistic conservative fluid dynamics codes. \emph{Monthly
Notices of the Royal Astronomical Society}, \emph{429}(4), 3533.
\url{https://doi.org/10.1093/mnras/sts632}

\leavevmode\hypertarget{ref-sharma_electron_2007}{}%
Sharma, P., Quataert, E., Hammett, G. W., \& Stone, J. M. (2007).
Electron {Heating} in {Hot} {Accretion} {Flows}. \emph{The Astrophysical
Journal}, \emph{667}, 714--723. \url{https://doi.org/10.1086/520800}

\leavevmode\hypertarget{ref-stone_athena_2020}{}%
Stone, J. M., Tomida, K., White, C. J., \& Felker, K. G. (2020). The
{Athena}++ {Adaptive} {Mesh} {Refinement} {Framework}: {Design} and
{Magnetohydrodynamic} {Solvers}. \emph{The Astrophysical Journal
Supplement Series}, \emph{249}, 4.
\url{https://doi.org/10.3847/1538-4365/ab929b}

\leavevmode\hypertarget{ref-tiede_variational_2020}{}%
Tiede, P., Broderick, A. E., \& Palumbo, D. C. M. (2020). Variational
{Image} {Feature} {Extraction} for the {EHT}. \emph{arXiv:2012.07889
{[}Astro-Ph{]}}. \url{http://arxiv.org/abs/2012.07889}

\leavevmode\hypertarget{ref-toth_b0_2000}{}%
Tóth, G. (2000). The ∇·{B}=0 {Constraint} in {Shock}-{Capturing}
{Magnetohydrodynamics} {Codes}. \emph{Journal of Computational Physics},
\emph{161}(2), 605--652. \url{https://doi.org/10.1006/jcph.2000.6519}

\leavevmode\hypertarget{ref-werner_non-thermal_2018}{}%
Werner, G. R., Uzdensky, D. A., Begelman, M. C., Cerutti, B., \&
Nalewajko, K. (2018). Non-thermal particle acceleration in collisionless
relativistic electron-proton reconnection. \emph{Monthly Notices of the
Royal Astronomical Society}, \emph{473}, 4840--4861.
\url{https://doi.org/10.1093/mnras/stx2530}

\leavevmode\hypertarget{ref-white_extension_2016}{}%
White, C. J., Stone, J. M., \& Gammie, C. F. (2016). An {Extension} of
the {Athena}++ {Code} {Framework} for {GRMHD} {Based} on {Advanced}
{Riemann} {Solvers} and {Staggered}-mesh {Constrained} {Transport}.
\emph{The Astrophysical Journal Supplement Series}, \emph{225}(2), 22.
\url{https://doi.org/10.3847/0067-0049/225/2/22}

\leavevmode\hypertarget{ref-Wielgus_monitoring_2020}{}%
Wielgus, M., Akiyama, K., Blackburn, L., Chan, C., Dexter, J., Doeleman,
S. S., Fish, V. L., Issaoun, S., Johnson, M. D., Krichbaum, T. P., Lu,
R.-S., Pesce, D. W., Wong, G. N., Bower, G. C., Broderick, A. E., Chael,
A., Chatterjee, K., Gammie, C. F., Georgiev, B., \ldots{} Zhu, Z.
(2020). {Monitoring the Morphology of M87* in 2009-2017 with the Event
Horizon Telescope}. \emph{901}(1), 67.
\url{https://doi.org/10.3847/1538-4357/abac0d}

\leavevmode\hypertarget{ref-zilhao_dynamic_2014}{}%
Zilhão, M., \& Noble, S. C. (2014). Dynamic fisheye grids for binary
black hole simulations. \emph{Classical and Quantum Gravity},
\emph{31}(6), 065013.
\url{https://doi.org/10.1088/0264-9381/31/6/065013}

\end{CSLReferences}

\end{document}